%% file: draft.tex
\documentclass[12pt,letterpaper]{article}

\usepackage[utf8]{inputenc}
\usepackage[margin=1in]{geometry}
\usepackage{amsmath}
\usepackage{amsfonts}
\usepackage{amssymb}
\usepackage{amsthm}
\usepackage{graphicx}
\usepackage{hyperref}
\usepackage{natbib}
\usepackage{booktabs}
\usepackage{longtable}
\usepackage{array}
\usepackage{setspace}
\usepackage{float}
\usepackage[T1]{fontenc}
\usepackage{lmodern}
\usepackage{subcaption}

\graphicspath{{images/}}

\usepackage{array}
\renewcommand{\arraystretch}{0.85}
\title{\Large \textbf{Generative AI and the Reallocation of Time: Productivity, Leisure, and Fulfilling Work}\thanks{\protect\linespread{1}\protect\selectfont We thank Minjeong Kim for excellent assistance with the survey. This work was funded by the Bank of Korea. The views expressed herein are those of the authors and not necessarily those of the Bank of Korea.}}

\author{
Donghyun Suh\thanks{Bank of Korea. Email: donghyunsuh.econ@gmail.com.} \and 
Samil Oh\thanks{Bank of Korea. Email: samil.oh@bok.or.kr.}
}

\date{February 12, 2026}

\begin{document}

\maketitle

\begin{abstract}
\singlespacing
Using a representative survey of Korean workers, we provide evidence on the adoption of Generative AI (GenAI) and how GenAI reallocates time at work. We find that 51.8\% of workers use GenAI for work and GenAI  reduces working time by 3.8\%. However, these gains may not materialize in aggregate productivity statistics yet: the correlation between time savings and output changes is near zero. We show this disconnect arises because workers capture efficiency gains primarily as on-the-job leisure, rather than increasing their output. These findings suggest that standard productivity measures may understate AI’s impact by missing non-pecuniary welfare channels.
\end{abstract}


\newpage

\section{Introduction}

Generative AI (GenAI) represents a general purpose technology with the potential to significantly increase aggregate productivity; yet, the extent to which these efficiency gains will materialize as output growth remains uncertain. In this paper, we investigate these gains and their transmission using a representative survey of Korean workers. We document a divergence between efficiency gains and realized output: while GenAI reduces working time, we find a near-zero correlation between these time savings and output gains. Instead, time savings are primarily absorbed by non-pecuniary channels—specifically on-the-job leisure and task reallocation—implying that standard productivity statistics may understate the technology's welfare impact.

We make three contributions. First, we measure reported output changes alongside time savings due to GenAI. This design allows us to decompose the effects of GenAI into productivity and on-the-job leisure. The survey reveals that the correlation between time savings and output changes is near zero (0.008). This result implies that time savings overstate the economy-wide output gains. Instead, the efficiency gains accrue to workers as increased on-the-job leisure rather than as higher measured output.

Second, we track how GenAI reallocates time between fulfilling and non-fulfilling tasks, capturing welfare effects through task composition. We assume that workers have preferences over tasks; technologies that shift time toward fulfilling activities can raise welfare even in the absence of productivity gains. This channel establishes a link between task composition and worker welfare: if AI automates fulfilling work, the effects of efficiency gains on worker welfare may be muted as workers end up spending more time on non-fulfilling work.

Third, we connect time savings, output changes, and task reallocation to overall job satisfaction to provide a comprehensive welfare assessment. We find that greater satisfaction is associated with time savings even when output is flat, suggesting the dominant role of on-the-job leisure. However, relatively few workers report unambiguous greater satisfaction when judged jointly on time and task dimensions, revealing that GenAI's welfare impact remains highly heterogeneous at current adoption levels.

These findings suggest that the lack of aggregate productivity growth in early adoption phases can be evidence of a measurement artifact: the gains are real but ``latent'' in the form of reduced hours and altered task composition. While welfare gains currently operate through non-pecuniary channels, realizing the productivity gains will likely require organizational changes that shift incentives from time-saving to output-expansion.

The remainder of this paper proceeds as follows. Section \ref{sec:data} describes our survey design and measurement approach. Section \ref{sec:adoption} documents adoption patterns and Section \ref{sec:time_savings} analyzes time effects. Section \ref{sec:disconnect} presents our core analysis of the time-output disconnect and process frictions, and develops a conceptual framework for interpreting these patterns. Section \ref{sec:welfare} analyzes the welfare channels—leisure, fulfilling tasks, and satisfaction. Section Section \ref{sec:conclusion} concludes.

\section{Data and Methodology} \label{sec:data}

\subsection{Survey Design and Sampling}

We fielded a nationally representative online survey of 5,512 Korean workers (employed individuals aged 15--64) between May 19 and June 17, 2025. The survey was conducted using the Korea Research web panel via Computer-Assisted Web Interviewing (CAWI) and Computer-Assisted Mobile Interviewing (CAMI). The sampling frame was based on the \textit{2024 Regional Employment Survey} conducted by Statistics Korea.

To ensure sufficient statistical power for analyzing heterogeneous task effects, we utilized a stratified sampling design with a three-step allocation scheme. First, we stratified the population by Occupation (9 major groups), Age (5 groups), and Gender. Second, to address the skewed distribution of employment across occupations, we applied a square-root proportional allocation across 53 medium-level occupation categories. This approach ensures that smaller occupational groups (e.g., skilled agricultural workers, machine operators) are sufficiently represented compared to large groups (e.g., office clerks), allowing for granular task-level analysis. Third, within these occupational quotas, we applied proportional allocation for age and gender.

To correct for the over-sampling of smaller occupations and any non-response bias, we apply post-stratification weights throughout the analysis. The weights are constructed in three stages: design weights (inverse of selection probability), non-response adjustment, and raking to match the marginal distributions of occupation, age, and gender in the 2024 population benchmarks. The resulting sampling error is $\pm$1.32 percentage points at the 95\% confidence level.

\subsection{Measurement of Key Variables}

Our survey instrument was designed to disentangle measuring time inputs from output outcomes. The questionnaire followed a ``bottom-up'' logic to minimize recall bias.

\paragraph{Time Allocation and Savings.}
We first asked respondents to report their \textit{total} weekly hours and, crucially, their \textit{actual} working hours (excluding breaks and idle time). Respondents then selected their primary tasks from a list of standardized activities customized to their occupation code (e.g., ``coding'' for IT workers, ``driving'' for transport workers). 

To measure time savings, we did not simply ask for a total estimate. Instead, we asked respondents to evaluate how GenAI affected the time required for their specific selected tasks. Respondents reported the change in weekly hours (e.g., ``reduced by 30 minutes per day over 5 days''). We measure the share of time saved as the ratio of the reported change to the actual work hours.

\paragraph{Output Changes.}
To measure the total output volume of each task, we asked: \textit{``If your individual output volume prior to using GenAI was 100, what is it now?''}. Respondents answered on a numeric scale (e.g., 80, 100, 120), allowing us to measure productivity changes even if working time remained constant. This allows us to distinguish between workers who use efficiency gains to produce more (output $>100$) versus those who use them to work less (output $\approx 100$, time savings $>0$).

\paragraph{Fulfilling Tasks.}
For each task performed, respondents rated whether the task provided ``personal fulfillment or a sense of achievement''. We aggregate these task-level ratings to construct worker-level indices of \textit{fulfilling work share}, allowing us to track whether GenAI substitutes for drudgery or meaningful activities.

\subsection{Adoption and Usage Definitions}

We distinguish four adoption measures. \emph{Overall ever-use} indicates whether a respondent has used GenAI in any context (63.5\%). \emph{Work-related ever-use} captures use specifically for job tasks (51.8\%). \emph{Used last week} (37.4\%) identifies active users who utilized GenAI in the week preceding the survey; this forms our primary sample for the intensive margin analysis. Finally, \emph{usage intensity} captures the average daily duration of GenAI use among adopters.

\subsection{Sample Characteristics}

Weighted descriptive statistics summarizing the sample composition appear in Table \ref{tab:sample_summary}. We observe balanced gender representation, an age distribution concentrated between 30 and 59, and relatively high educational attainment. The table also documents the headline adoption rates used throughout the analysis.

\begin{table}[htp]
\centering
\caption{Weighted sample summary statistics}
\label{tab:sample_summary}
\input{outputs/sample_summary.tex}
\end{table}

\subsection{Estimation}

All descriptive statistics and regression models employ the survey weights described in Section 2.1. For adoption analysis, we estimate weighted logistic regressions for extensive margins and ordinary least squares (OLS) for usage intensity. For the time and output analysis, we restrict the sample to users and estimate:
\begin{equation}
    Y_{i} = \alpha + \beta X_{i} + \gamma_{occ} + \gamma_{ind} + \gamma_{region} + \epsilon_{i}
\end{equation}
where $Y_{i}$ is the outcome (e.g., percentage change in hours, output index) for worker $i$, $X_{i}$ is a vector of worker characteristics (age, education, income, labor supply elasticity), and $\gamma$ represents fixed effects for occupation, industry, and region. Standard errors are robust to heteroskedasticity.

\section{Adoption Patterns of GenAI} \label{sec:adoption}

\subsection{Extensive Margin: Who Adopts GenAI?}

\begin{figure}[t]
\centering
\includegraphics[width=0.90\textwidth]{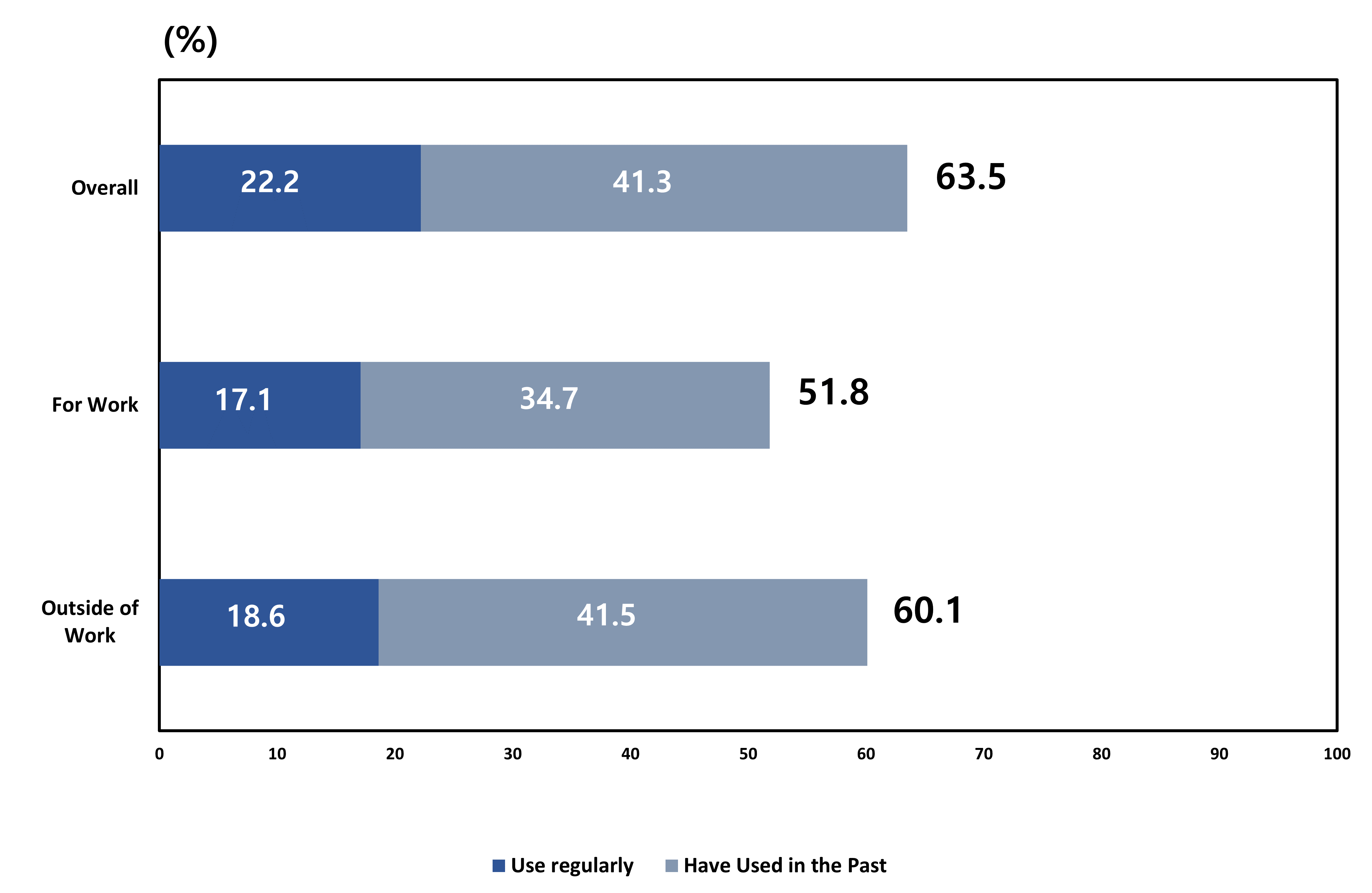}
\caption{Generative AI adoption rates among Korean workers}
\label{fig:adoption_overall}
\end{figure}

Generative AI has diffused rapidly into the Korean workplace. As shown in Figure \ref{fig:adoption_overall}, 63.5\% of workers report having ever used GenAI in any context. Crucially, workplace adoption is nearly as high: 51.8\% of workers report using GenAI specifically for work-related tasks. This suggests that GenAI is not merely a consumer novelty but has gained substantial traction as a production technology.

\begin{figure}[htp]
\centering
\begin{subfigure}{\textwidth}
  \centering
  \includegraphics[width=0.90\textwidth]{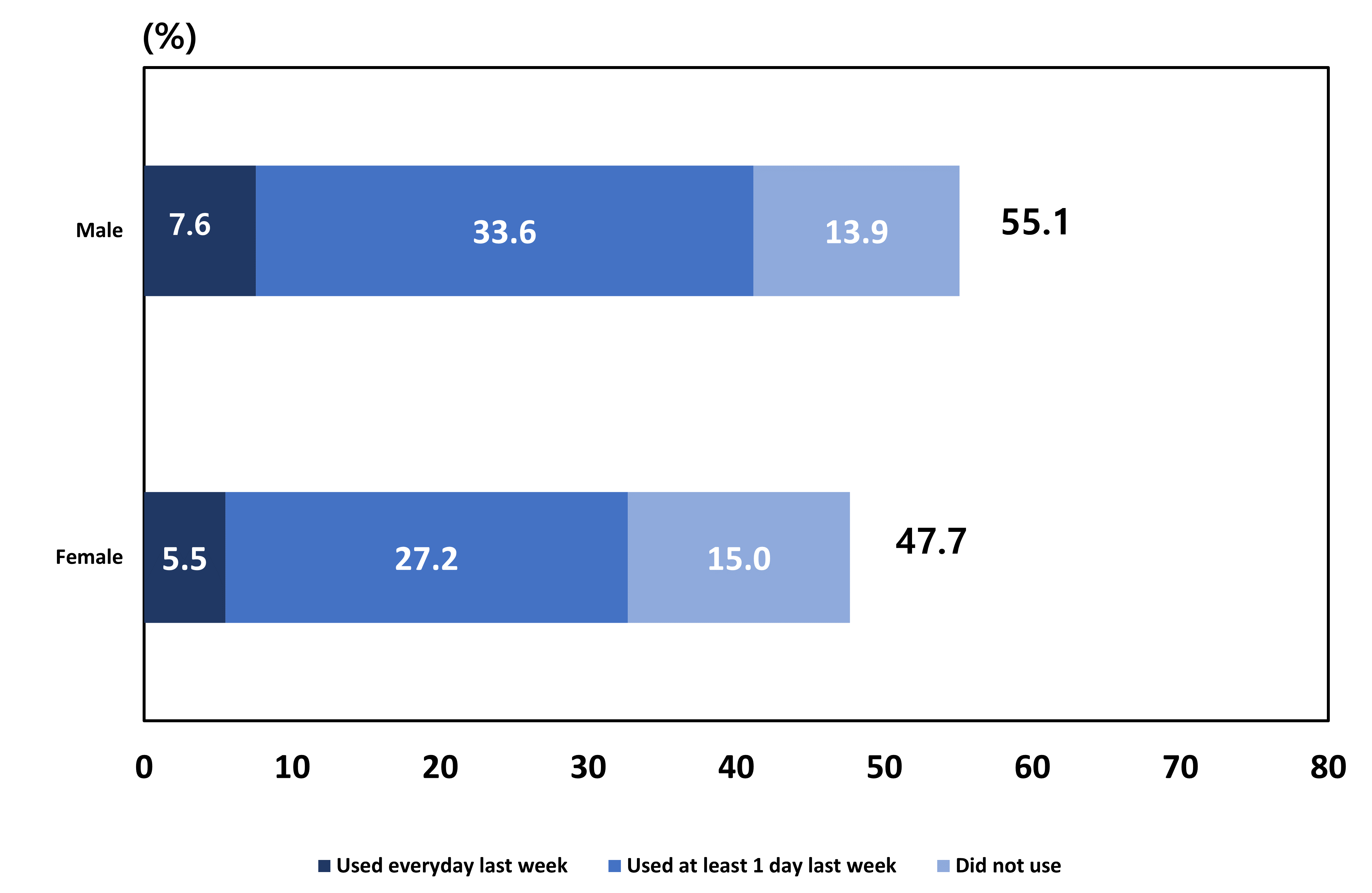}
  \caption{GenAI usage by gender}
  \label{fig:adoption_gender}
\end{subfigure}

\vspace{0.5cm}

\begin{subfigure}{\textwidth}
  \centering
  \includegraphics[width=0.90\textwidth]{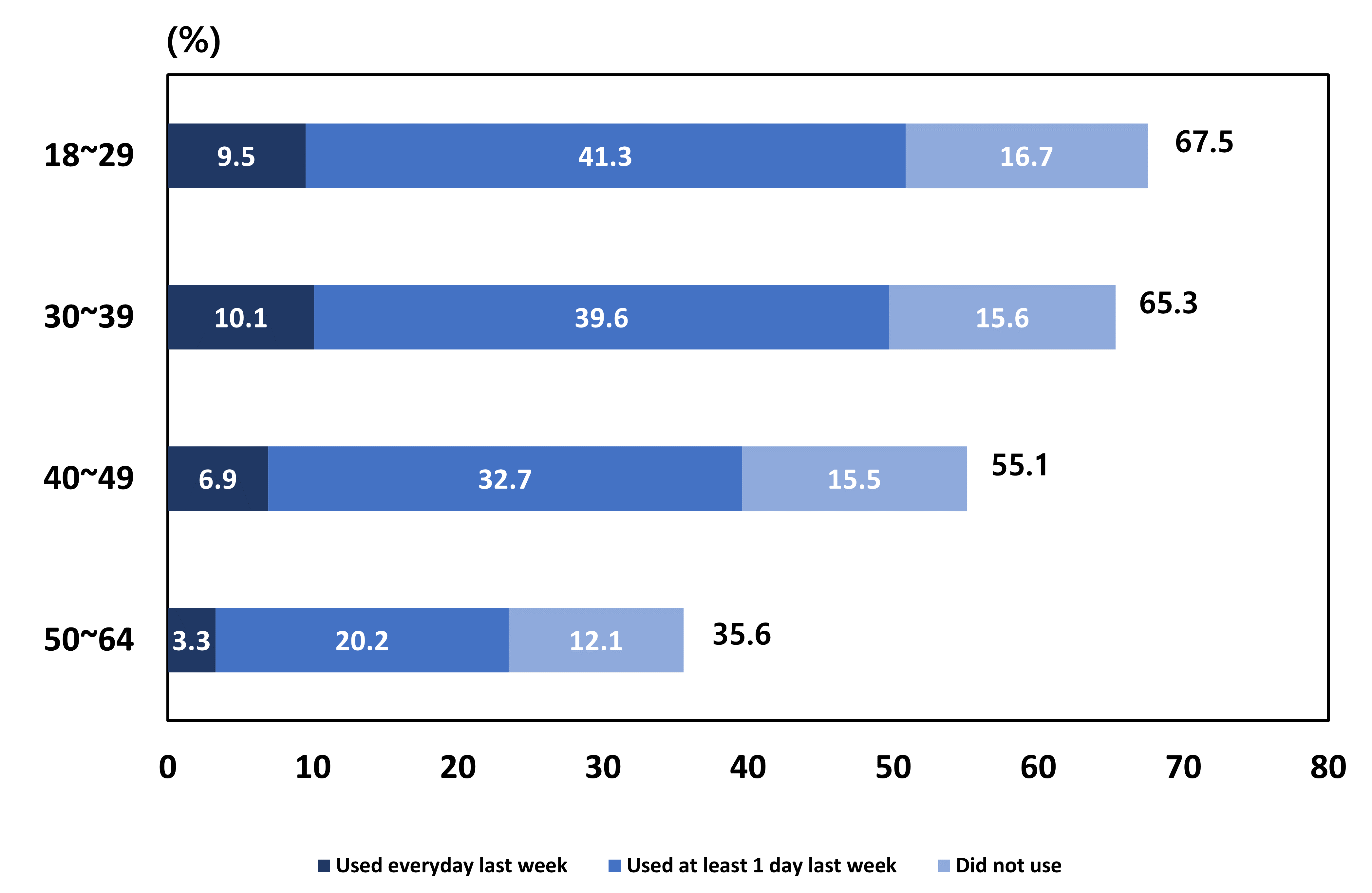}
  \caption{GenAI usage by age group}
  \label{fig:adoption_age}
\end{subfigure}

\caption{GenAI usage by gender and age}
\label{fig:adoption_demographics1}
\end{figure}

\begin{figure}[htp]
\centering
\begin{subfigure}{\textwidth}
  \centering
  \includegraphics[width=0.90\textwidth]{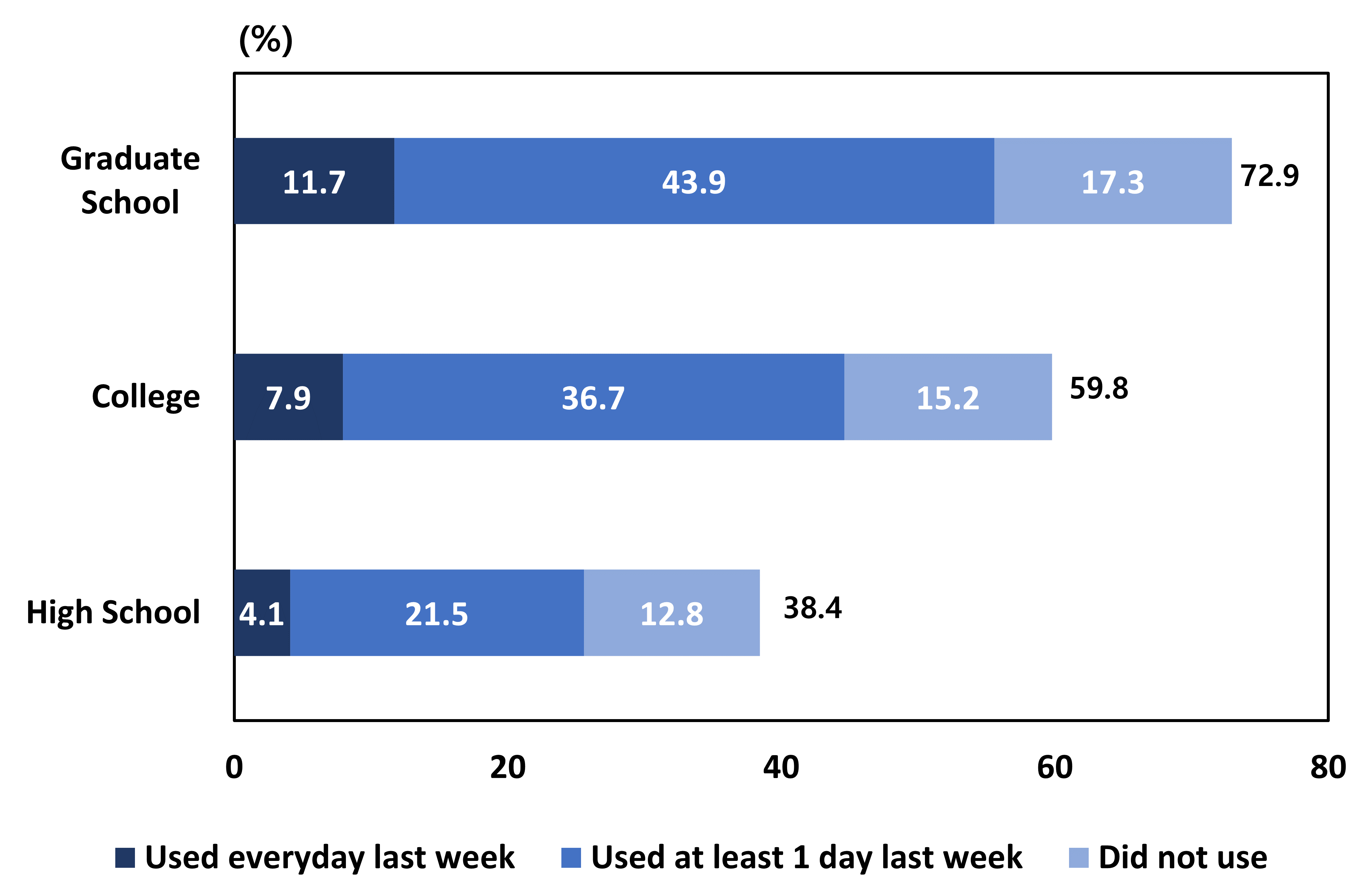}
  \caption{GenAI usage by education}
  \label{fig:adoption_edu}
\end{subfigure}

\vspace{0.5cm}

\begin{subfigure}{\textwidth}
  \centering
  \includegraphics[width=0.90\textwidth]{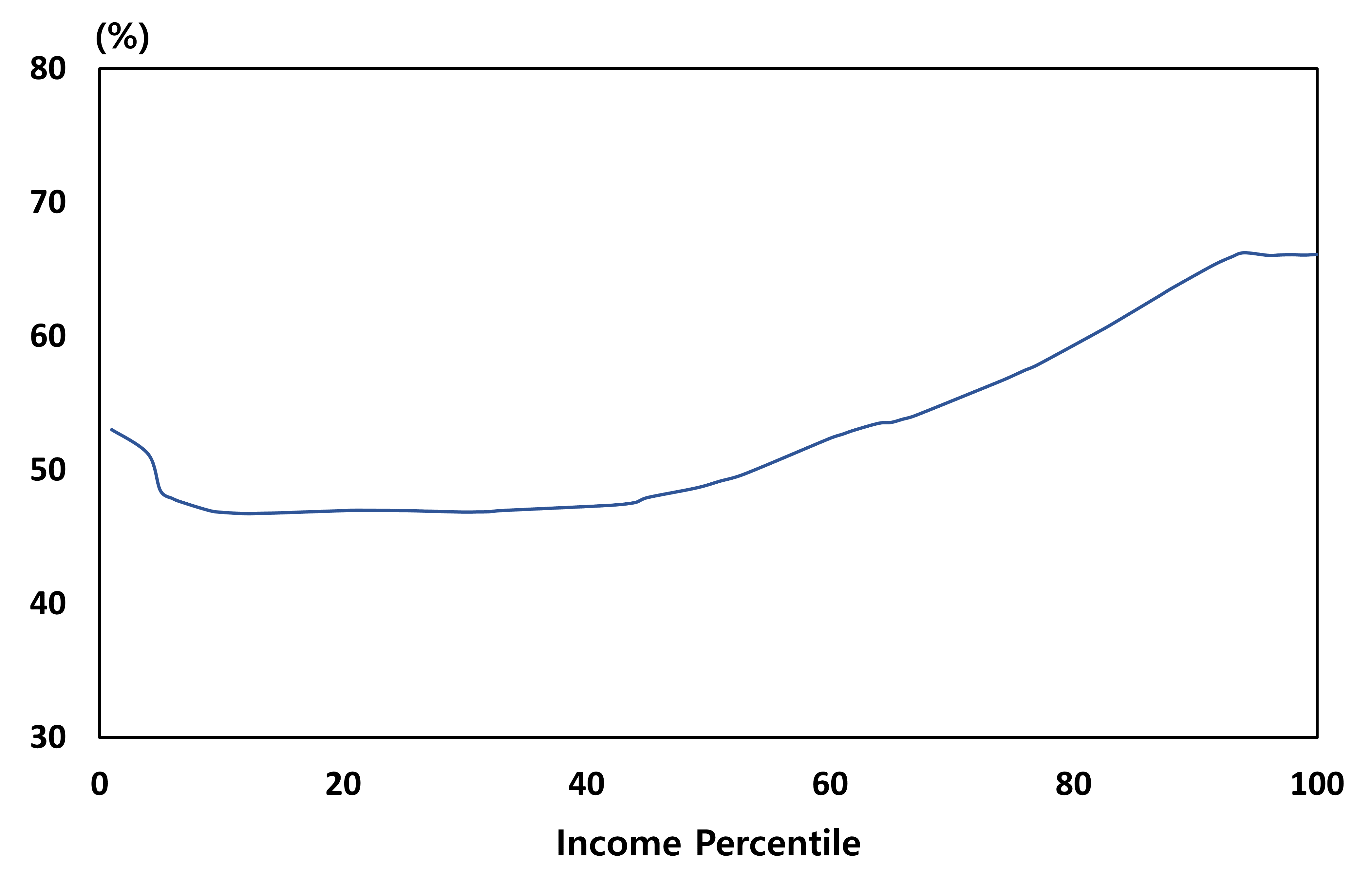}
  \caption{GenAI usage by income percentile}
  \label{fig:adoption_income}
\end{subfigure}

\caption{GenAI usage by education and income}
\label{fig:adoption_demographics2}
\end{figure}

Adoption exhibits steep gradients across demographic and socioeconomic lines. Figures \ref{fig:adoption_gender}--\ref{fig:adoption_edu} display participation rates for the ''Used last week'' sample (those using AI for work in the reference week). We observe three distinct patterns:
\begin{enumerate}
    \item \textbf{Age:} Adoption peaks among workers in their 20s and 30s and declines monotonically with age. This is consistent with the "digital native" hypothesis, where younger cohorts face lower costs of learning new interfaces.
    \item \textbf{Education:} There is a sharp educational divide. Workers with a four-year college degree or higher adopt at significantly higher rates than those with high school education or less. This aligns with the view of GenAI as a skill-biased technology in the early adoption phase.
    \item \textbf{Income:} Adoption is positively correlated with income (Figure \ref{fig:adoption_income}), likely reflecting the concentration of high-income earners in cognitive, white-collar occupations where GenAI is most applicable.
\end{enumerate}

\begin{figure}[htp]
\centering
\begin{subfigure}{\textwidth}
  \centering
  \includegraphics[width=0.90\textwidth]{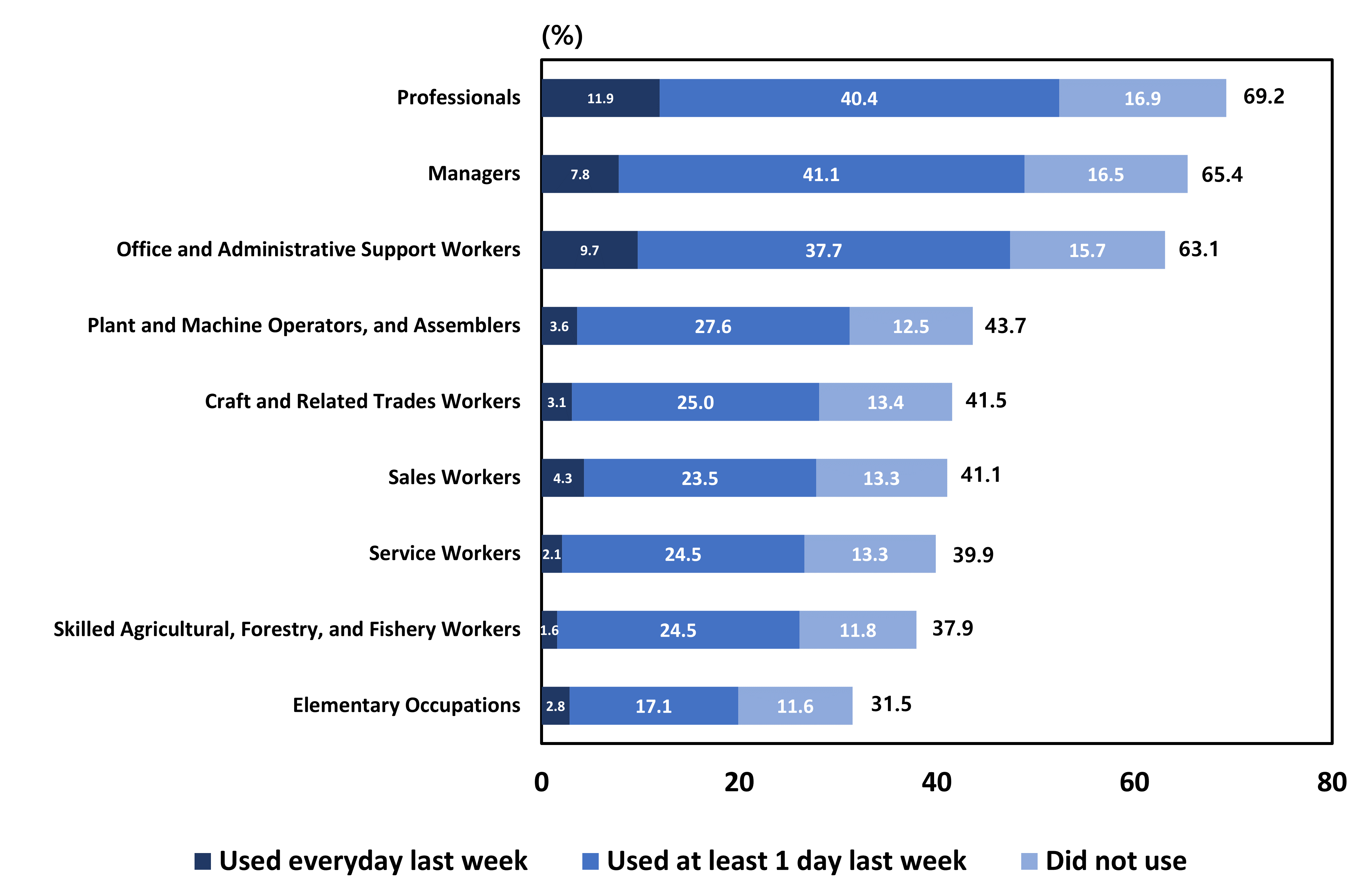}
  \caption{GenAI usage by occupation}
  \label{fig:adoption_occ}
\end{subfigure}

\vspace{0.5cm}

\begin{subfigure}{\textwidth}
  \centering
  \includegraphics[width=0.90\textwidth]{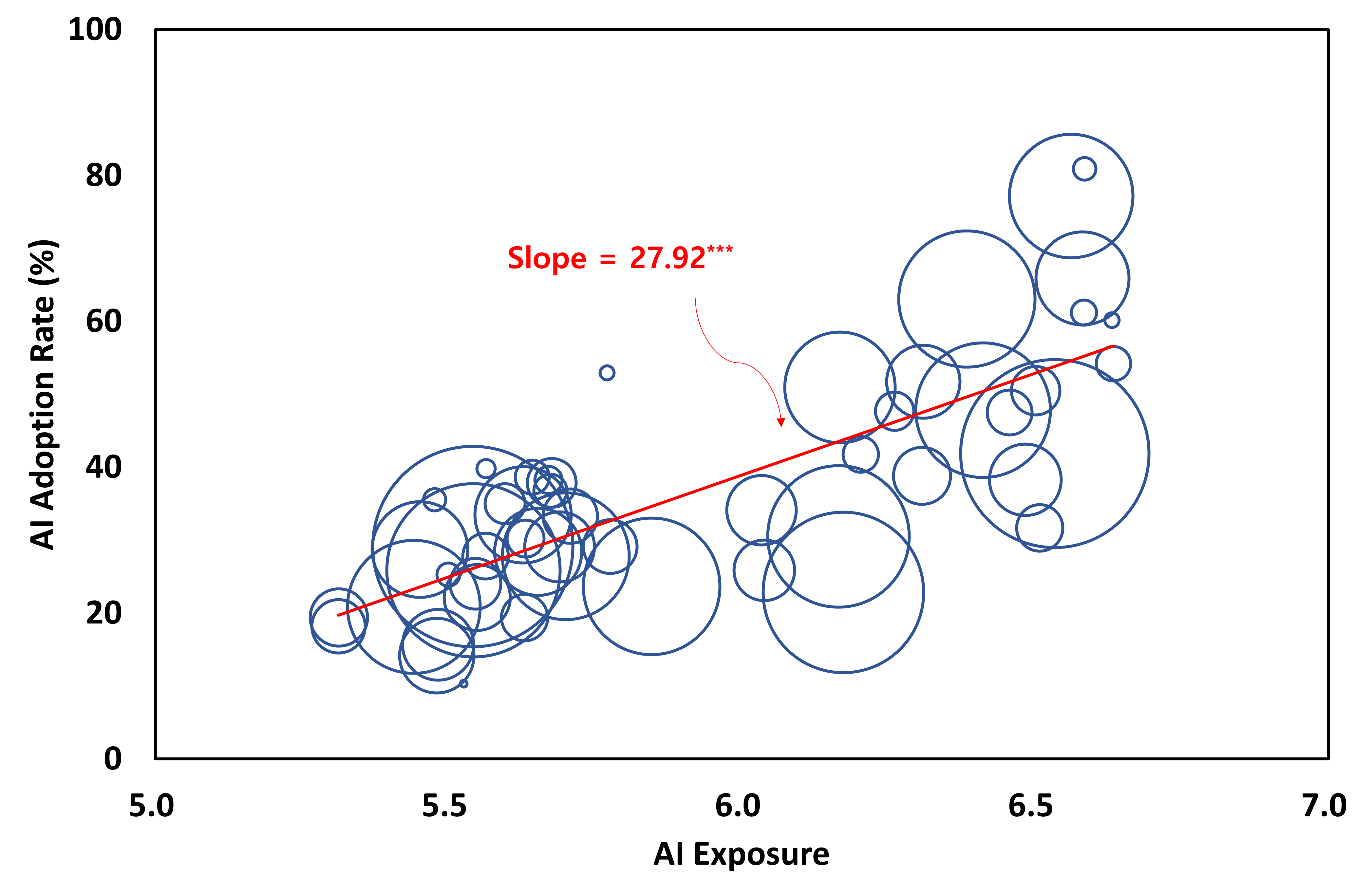}
  \caption{AI exposure and GenAI adoption across occupations}
  \label{fig:exposure_adoption}
\end{subfigure}

\caption{GenAI adoption patterns across occupations}
\label{fig:occupation_analysis}
\end{figure}

Occupational heterogeneity is profound. As shown in Figure \ref{fig:adoption_occ}, adoption is concentrated in Professional, Managerial, and Office sectors—occupations traditionally viewed as "cognitive" and "non-routine." In contrast, Service workers, Sales workers, and Manual laborers show markedly lower adoption rates. 

Figure \ref{fig:exposure_adoption} confirms this relationship:

We further examine whether these occupational adoption rates align with the predictions of AI exposure. Figure \ref{fig:exposure_adoption} plots the adoption rate for each occupation against its AI Exposure Index, constructed following the methodology of \cite{felten2021occupational}. Occupations with higher scores on the AI Exposure Index exhibit systematically higher realized adoption rates.

\subsection{Intensive Margin: Frequency and Duration}

\begin{figure}[htp]
\centering
\includegraphics[width=0.90\textwidth]{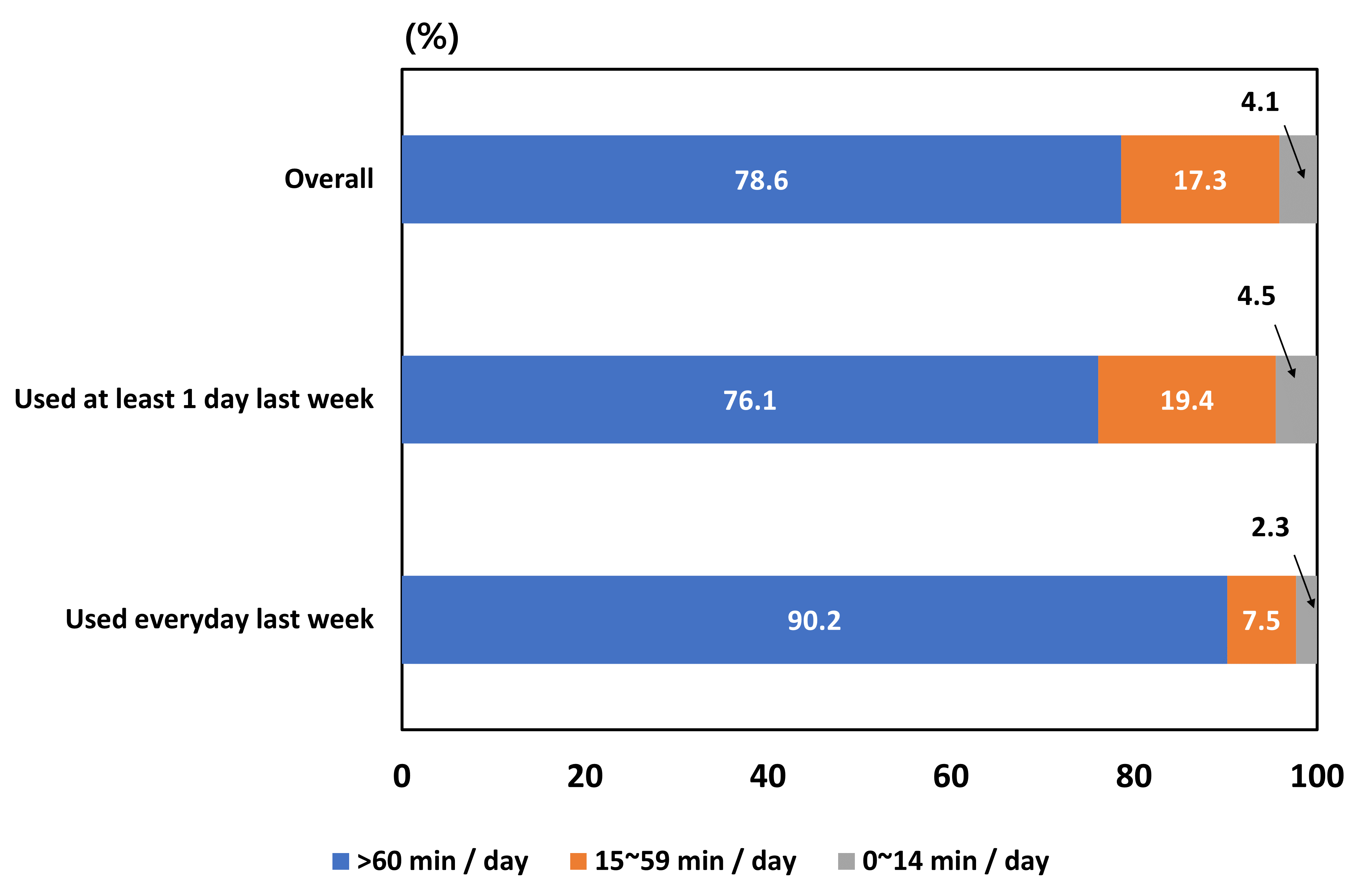}
\caption{Distribution of daily GenAI usage among adopters}
\label{fig:adoption_intensity}
\end{figure}

Conditional on adoption, usage intensity varies widely. The distribution of self-reported daily usage hours (Figure \ref{fig:adoption_intensity}) reveals that for a majority of users, GenAI assists their work for a significant amount of time. The results suggest that GenAI is already incorporated into the work process for a large fraction of workers.

\subsection{Determinants of Adoption}

To formalize these patterns, we estimate simple regressions summarizing the drivers of adoption (Table \ref{tab:genai_main_results}). The results confirm that education, work effort (hours, motivation), and occupational context are significantly correlated with adoption. 

Notably, gender gaps persist even after controlling for occupation, income, and education, suggesting that unobservable factors or differential exposure to technology within job titles may play a role. Furthermore, intrinsic work motivation is a strong predictor of adoption. One interpretation is that early adopters are often those seeking to augment their productivity.

\input{outputs/genai_main_results.tex}

\section{Time-saving Effects of GenAI} \label{sec:time_savings}

Now we turn to the productivity effect of GenAI. To do so, we start by analyzing the amount of time savings due to GenAI.

\subsection{Magnitude and Distribution of Time Savings}

We find that among workers who used GenAI in the week preceding the survey, working time decreased by an average of 3.8\%. For the overall workforce, including non-users, this translates to a 1.4\% reduction in average active hours at work.

\begin{figure}[htp]
\centering
\includegraphics[width=0.90\textwidth]{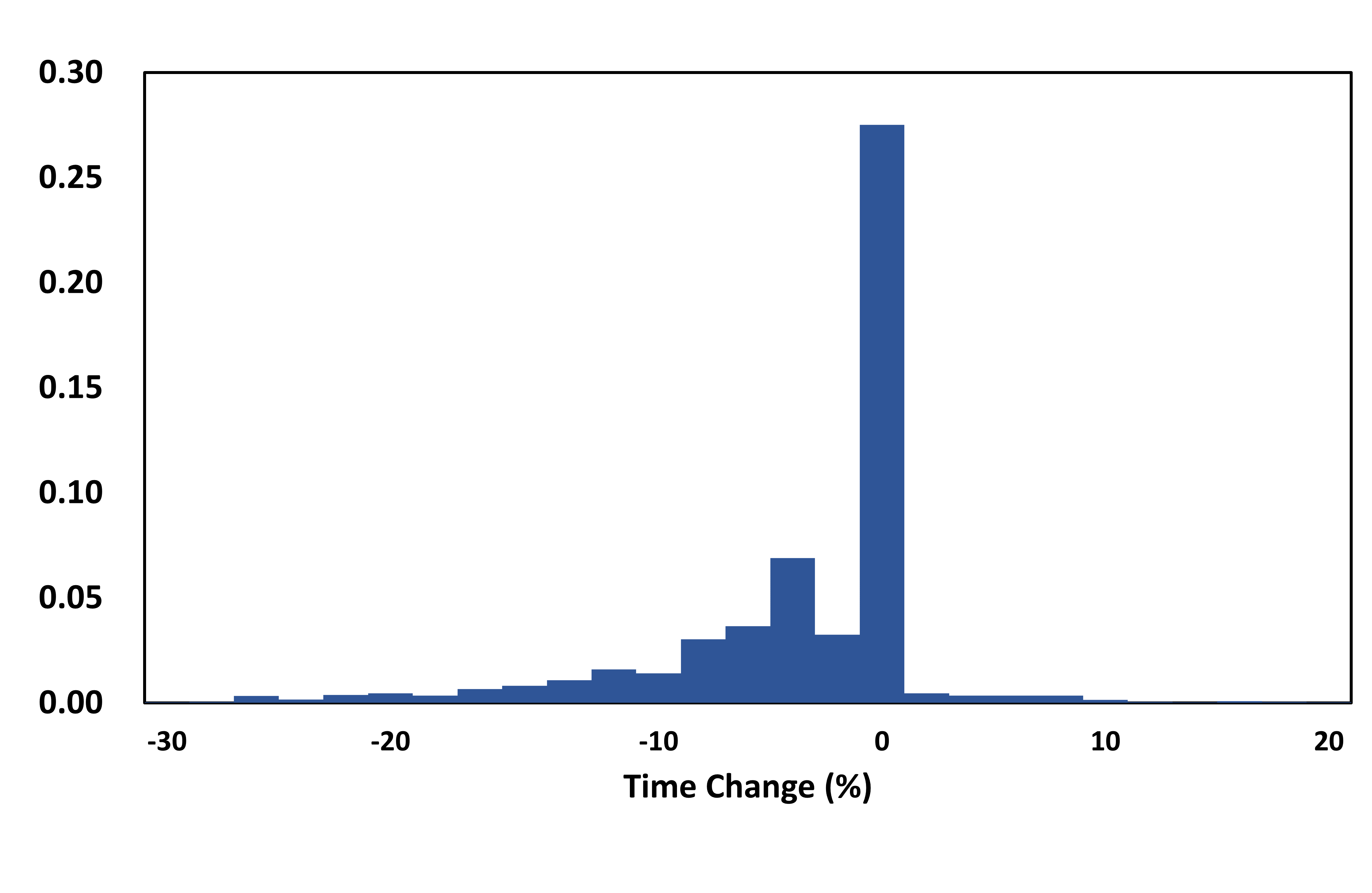}
\caption{Distribution of percentage time savings among GenAI users}
\label{fig:time_histogram}
\end{figure}

Figure \ref{fig:time_histogram} displays the distribution of percentage time savings among users. Two features of this distribution are particularly notable. First, the mass of the distribution is concentrated near zero, indicating that for a large share of users (approximately 50\%), the technology has not yet fundamentally altered the time required for daily workflows. This suggests that simply ``adopting'' the tool does not guarantee immediate efficiency gains.

Second, the distribution exhibits a distinct left skew, revealing a subset of ``super-savers'' who report time reductions exceeding 20\%. This left tail suggests that for specific tasks or workers, the technology acts as a strong substitute for labor time. Conversely, there is a small but non-negligible right tail of users who report time \textit{increases}. As we discuss in Section \ref{sec:disconnect}, this likely reflects learning costs, prompting overhead, or coordination frictions associated with integrating a new technology into existing processes.

\subsection{Heterogeneity by Occupation and Exposure}

\begin{figure}[htp]
\centering
\includegraphics[width=0.90\textwidth]{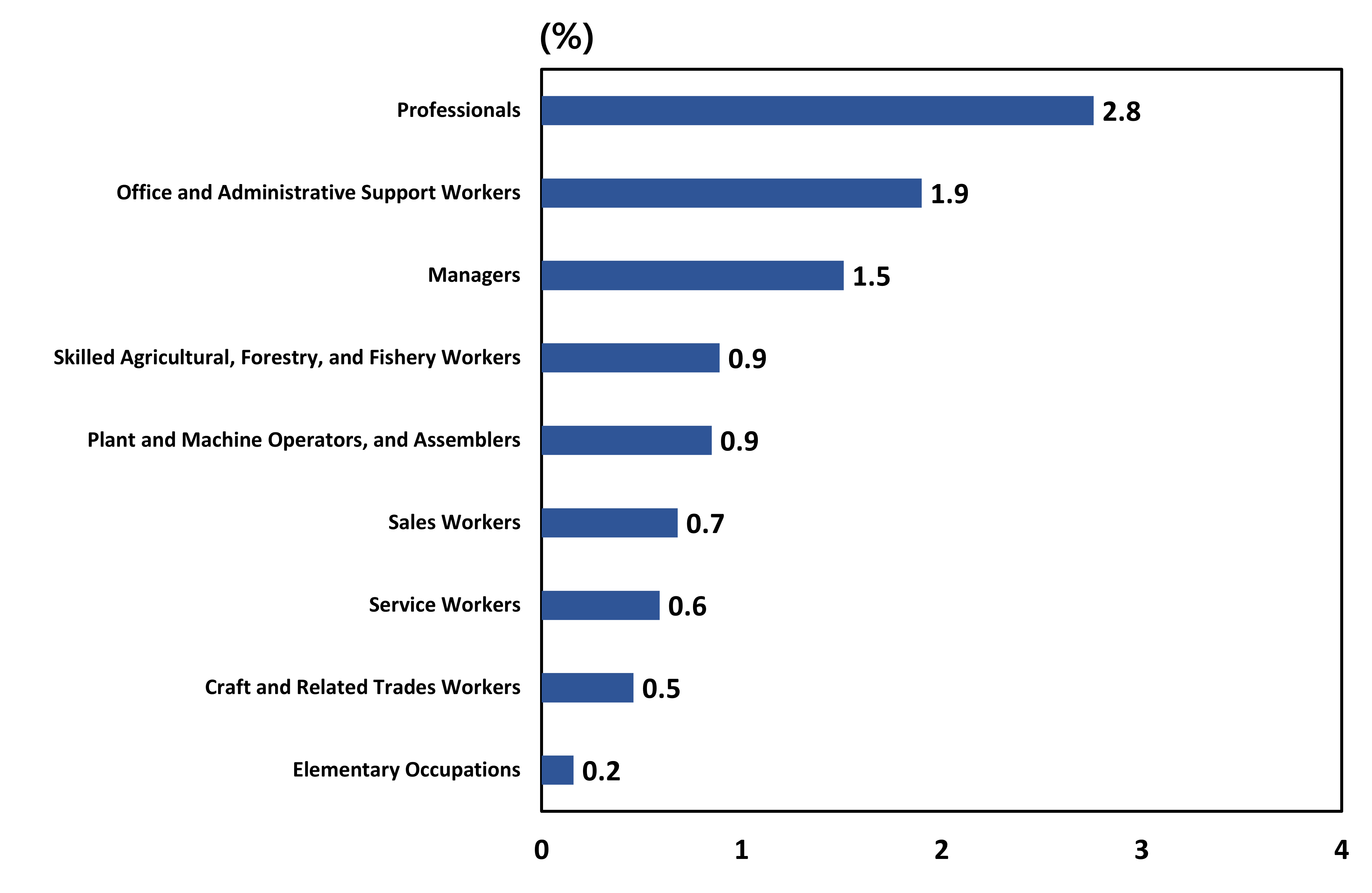}
\caption{Time savings by occupation}
\label{fig:time_occ}
\end{figure}

The time-saving effects of GenAI are unevenly distributed across the labor market. Figure \ref{fig:time_occ} presents the average percentage reduction in working hours by major occupation group. Time savings are most pronounced in professional and administrative occupations, where users report reductions of approximately 2.8\% and 1.9\%, respectively. In contrast, workers in manual, service, and elementary occupations report the smallest average time savings. This pattern is consistent with the view that GenAI primarily augments cognitive, information-intensive tasks rather than physical or interpersonal ones.

\begin{figure}[htp]
\centering
\begin{subfigure}{\textwidth}
  \centering
  \includegraphics[width=0.90\textwidth]{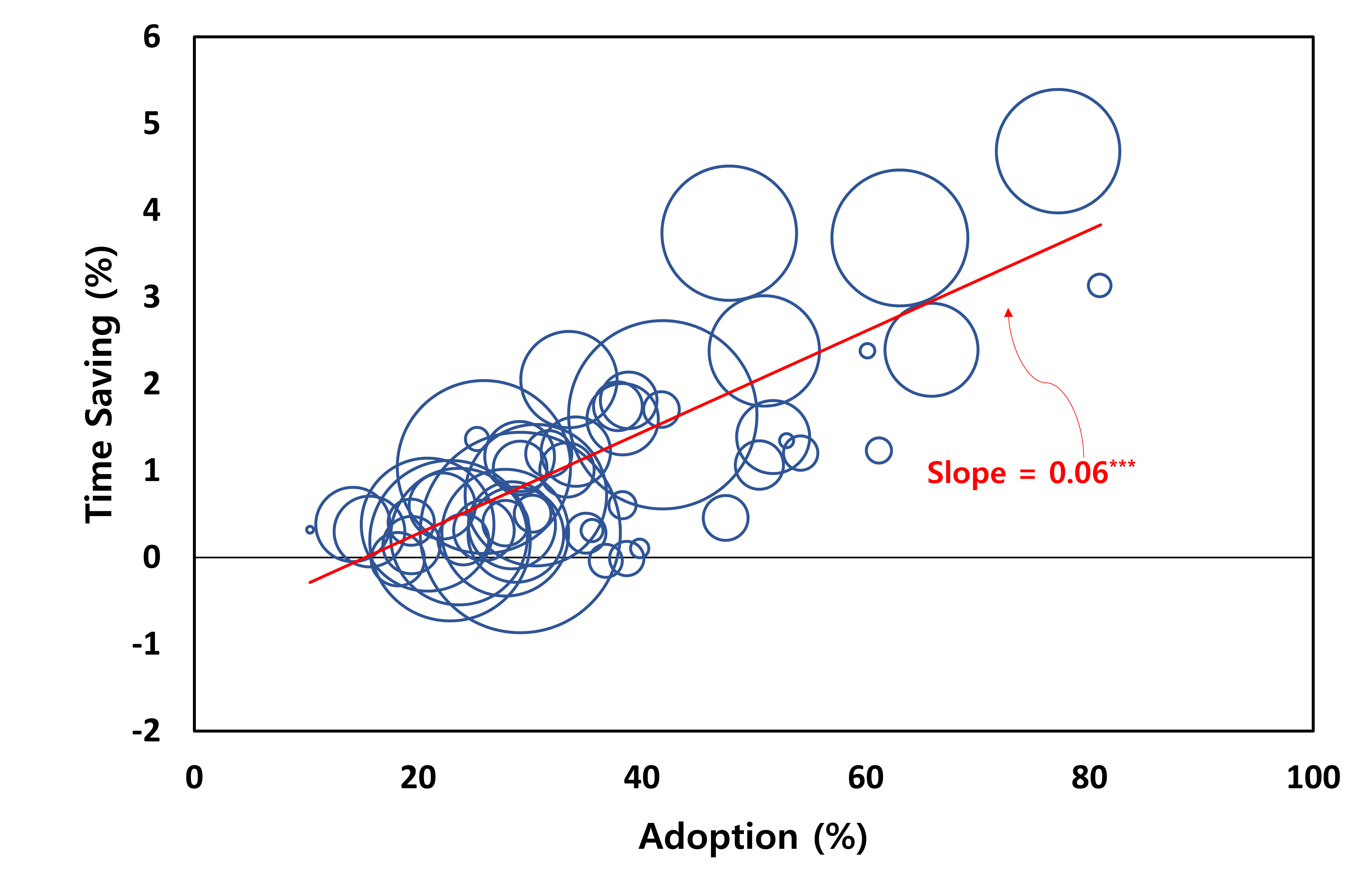}
  \caption{Adoption and time savings across occupations}
  \label{fig:time_adoption_scatter}
\end{subfigure}

\vspace{0.5cm}

\begin{subfigure}{\textwidth}
  \centering
  \includegraphics[width=0.90\textwidth]{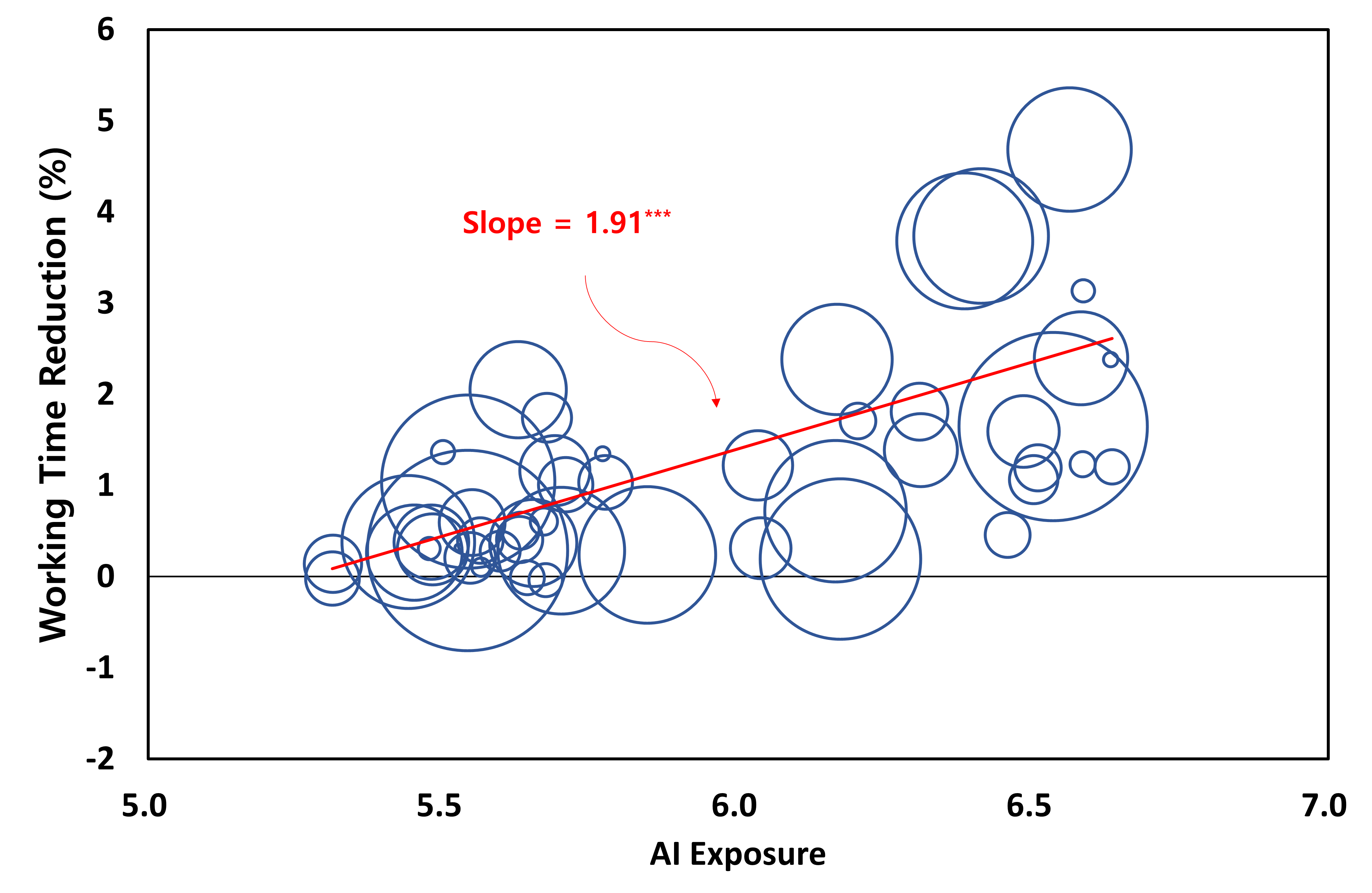}
  \caption{AI exposure and time savings across occupations}
  \label{fig:exposure_time_scatter}
\end{subfigure}

\caption{Relationship between adoption, AI exposure, and time savings}
\label{fig:time_savings_analysis}
\end{figure}

We further examine whether these realized savings align with the predictions of AI exposure. Figure \ref{fig:exposure_time_scatter} plots the average time reduction for each occupation against its AI Exposure Index, similarly with Figure \ref{fig:exposure_adoption}. We observe a positive correlation: occupations with higher exposure to language modeling technologies (e.g., analysts, programmers, administrative support) realize systematically larger reductions in working time.

Furthermore, there is a strong relationship between the extensive and intensive margins of the technology. Figure \ref{fig:time_adoption_scatter} shows that occupations with higher adoption rates also experience greater average time savings among users. This positive slope suggests that in sectors where the technology is widely applicable, it also delivers greater efficiency gains per user.

\subsection{Task-Level Mechanisms}

\begin{figure}[htp]
\centering
\includegraphics[width=0.90\textwidth]{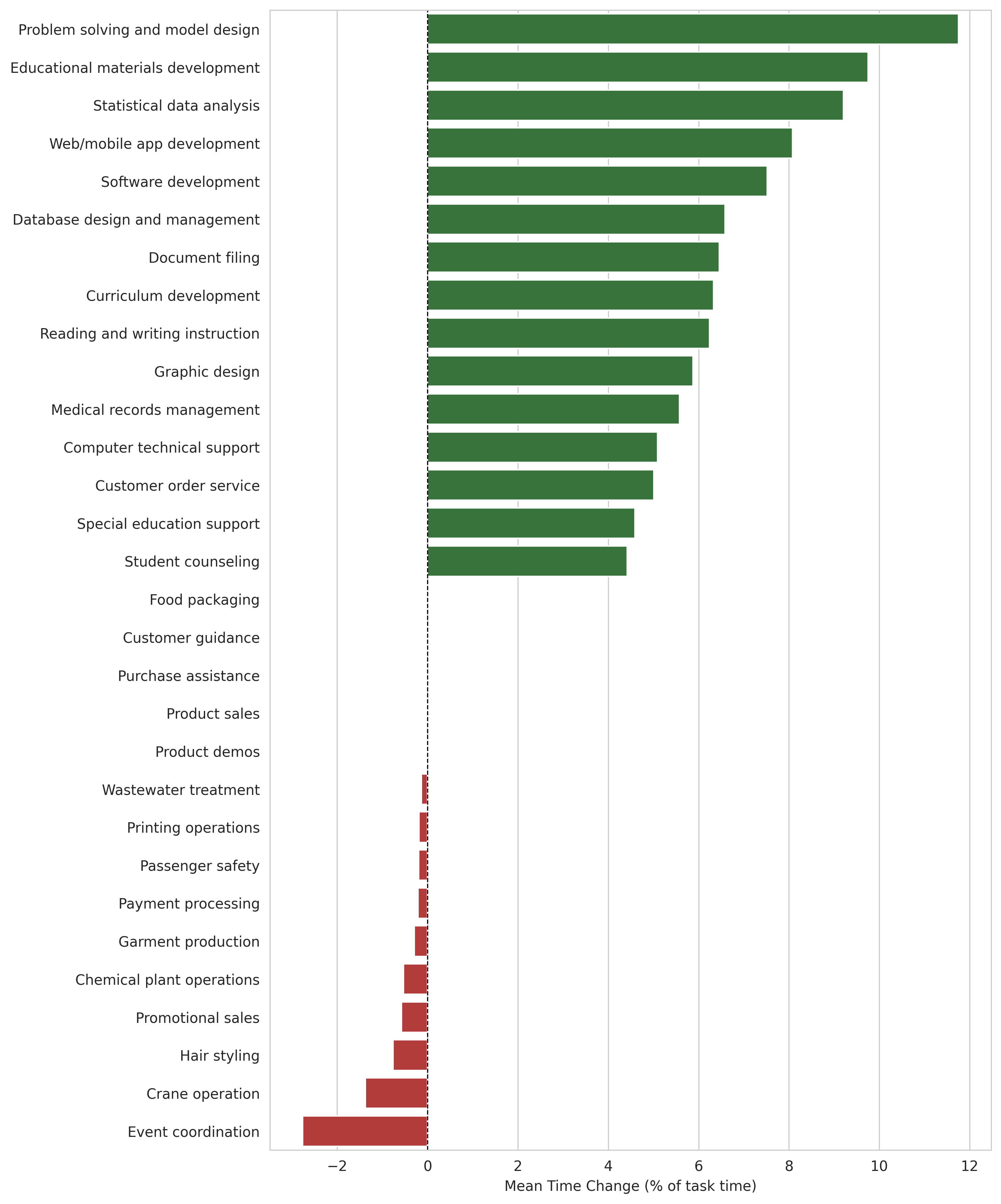}
\caption{Time savings for specific tasks (top 15 and bottom 15 among tasks with $\geq$30 respondents)}
\label{fig:time_task}
\end{figure}

To understand the source of these savings, we decompose the effect by task type. Figure \ref{fig:time_task} reports the average time savings for specific work activities among tasks chosen by at least 30 respondents (156 tasks total). The results reveal a clear dichotomy in GenAI's capability based on the nature of the task.

Tasks involving information processing and generation—specifically coding, data analysis, and document drafting—experience the largest time reductions. The top time-saving tasks include problem-solving and data analysis (11.8\% reduction) and educational materials development (9.8\%). Among all analyzed tasks, 140 (90\%) show positive time savings, with a mean reduction of 1.9\% across all tasks.

In contrast, a small subset of tasks (10 tasks, 6\%) show time increases, primarily involving physical coordination and customer complaint handling. This suggests that while GenAI is an effective substitute for cognitive labor, it may add frictions to tasks requiring high-context human judgment or physical presence.

To formalize these determinants, the second column of Table \ref{tab:genai_main_results} presents the estimation results where the dependent variable is the percentage change in hours. Unlike the results on adoption, most control variables of interest lose statistical significance. 

Notably, experience is significantly and negatively correlated with the amount of time saved. In other words, workers with less experience save more time by using GenAI. The results indicate that a 20-year difference in experience is associated with a 1.92 percentage point gap in time savings, favoring less experienced workers.


\section{The Time-Output Disconnect} \label{sec:disconnect}

Having established that GenAI generates substantial time savings, we now turn to the question: do these time savings translate into more output? To test whether time savings translate into actual output gains, we asked respondents to estimate their task-level output volume relative to a pre-GenAI baseline of 100 (e.g., a response of 120 indicates a 20\% increase in output).

\subsection{Correlation between Time Savings and Output Changes}

\begin{figure}[htp]
\centering
\includegraphics[width=0.90\textwidth]{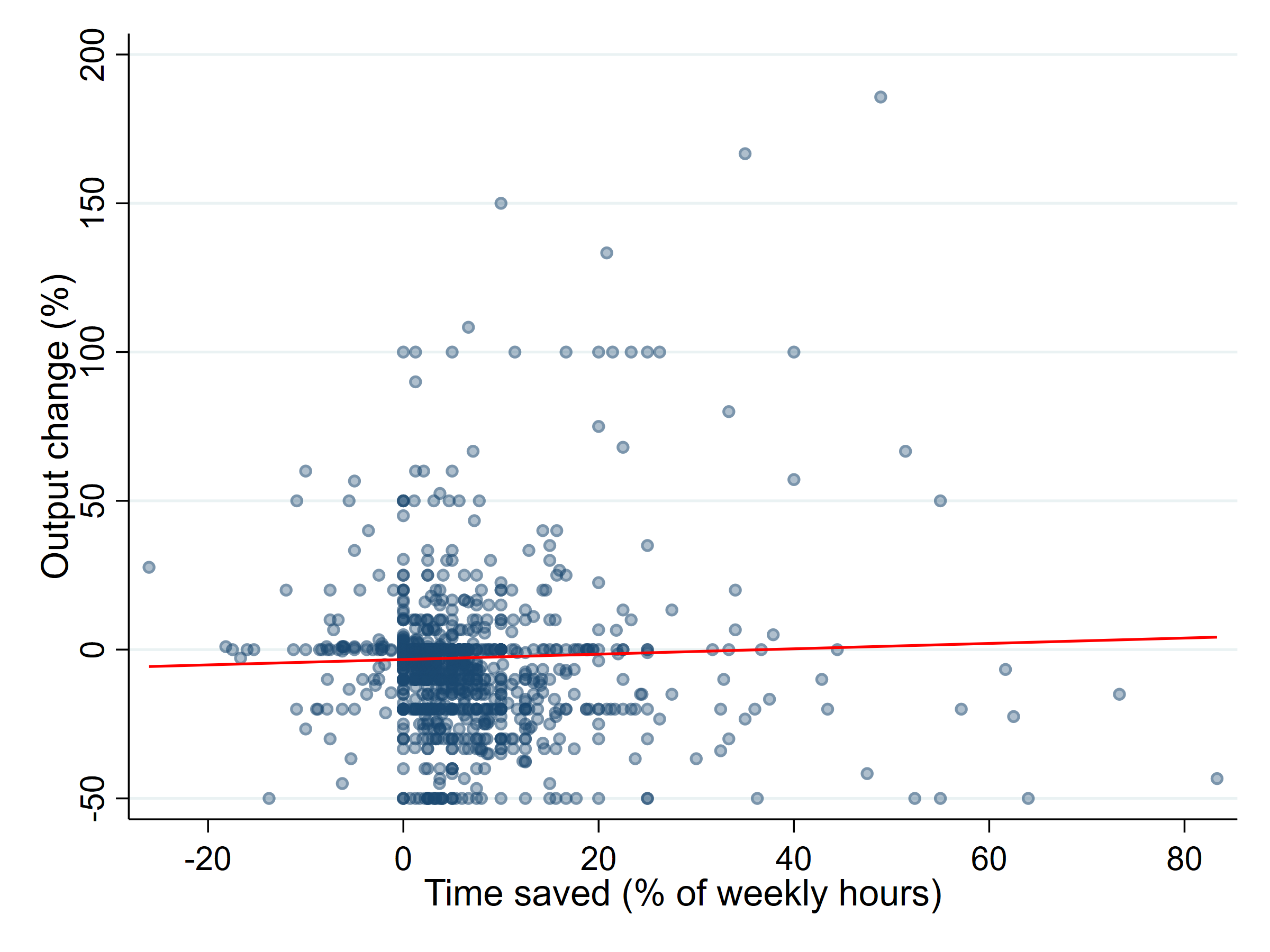}
\caption{Output change vs. time savings among users}
\label{fig:output_vs_time}
\end{figure}

Figure \ref{fig:output_vs_time} plots the relationship between percentage changes in working time and percentage changes in output volume among users. The scatter plot reveals no positive relationship between time savings and output gains. 

The weighted correlation between time savings and output changes is very low at 0.008. There is no systematic linear relationship where ``more time saved'' leads to ``more output produced''. While a cluster of workers appears at the origin, substantial mass lies off the diagonal.

Regression results in Table \ref{tab:output_regressions} confirm this lack of association. Even after conditioning on detailed demographic controls, occupation fixed effects, and industry fixed effects, the coefficient on time savings is statistically indistinguishable from zero. Restricting the sample only to those who saved time yields a similarly insignificant result.

\input{outputs/output_regressions.tex}

To further investigate the lack of a positive relationship between output and time savings, it is informative to examine the reasons for output \textit{declines}, reported by the respondents. Among those who report output declines (36.1\% of users), 40.4\% reported they are not used to using GenAI and 59.6\%. These results suggest that workers are still at an early phase in their adoption of GenAI. Furthermore, as workers become more experienced with the tools, output gains are likely to materialize. 





\section{Implications of GenAI for Worker Welfare} \label{sec:welfare}

If the efficiency gains from GenAI are not materializing as productivity gains, they must be absorbed elsewhere. We identify two primary channels in addition to the productivity effect through which workers capture the value of the technological shock: increased leisure and improved task composition.

To see how workers' satisfaction at work change with GenAI usage, we asked respondents about their change in work satisfaction due to GenAI.

\subsection{The Leisure Channel}

The most direct mechanism for welfare improvement is the conversion of time savings into on-the-job leisure. This represents a reduction in work intensity—completing the same bundle of tasks with less effort and time, thereby reducing the disutility of labor.

\begin{figure}[htp]
\centering
\includegraphics[width=0.72\textwidth]{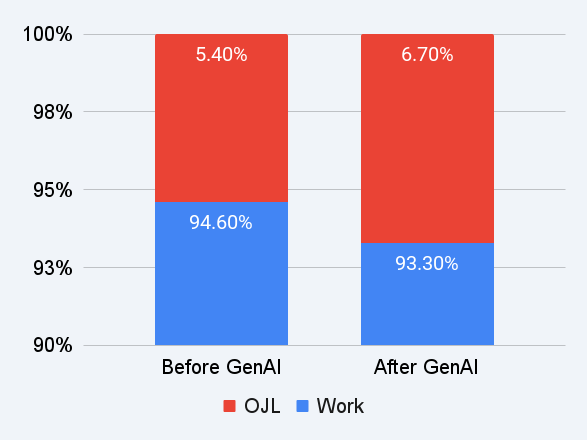}
\caption{Reallocation between work and on-the-job leisure}
\label{fig:OJL}
\end{figure}

Our survey allows us to decompose the total time at work into active work hours and on-the-job leisure. Figure \ref{fig:OJL} illustrates this decomposition. Among users, the efficiency shock enables a 1.3 percentage point increase in the share of time spent on on-the-job leisure.

Our results suggest that in the current institutional environment—where output quotas are often fixed or sticky—workers are rational agents who utilize the technology to reduce active work hours. An implication for measuring the effects of AI is that AI can increase worker welfare even if it is invisible in the standard productivity metrics.

\subsection{The Task Composition Channel}

The second mechanism for welfare improvement is the reallocation of time across different types of work. Even if aggregate output volume and work time remain constant, welfare improves if GenAI allows workers to substitute drudgery (routine, low-fulfillment tasks) for ``meaningful work'' (high-fulfillment tasks).

\paragraph{Heterogeneity in Time Savings.}
Time savings are highly concentrated. As shown in Table \ref{tab:task_concentration}, the top-saving task accounts for approximately 70\% of a user's total time reduction. This suggests that GenAI functions as a targeted tool for specific bottlenecks rather than a general-purpose accelerator of all work.


\input{outputs/task_concentration_tables}

\begin{figure}[htp]
\centering

\begin{subfigure}[b]{0.50\textwidth}
    \centering
    \includegraphics[width=\textwidth]{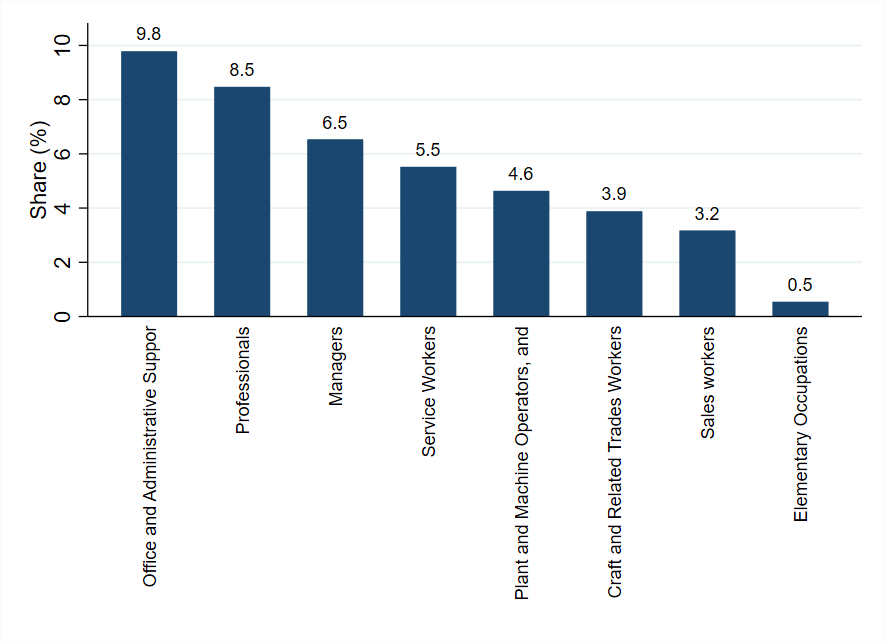}
    \caption{Occupations with increased fulfilling-task share}
    \label{fig:more_fulfill_occ}
\end{subfigure}

\vspace{0.1cm}

\begin{subfigure}[b]{0.50\textwidth}
    \centering
    \includegraphics[width=\textwidth]{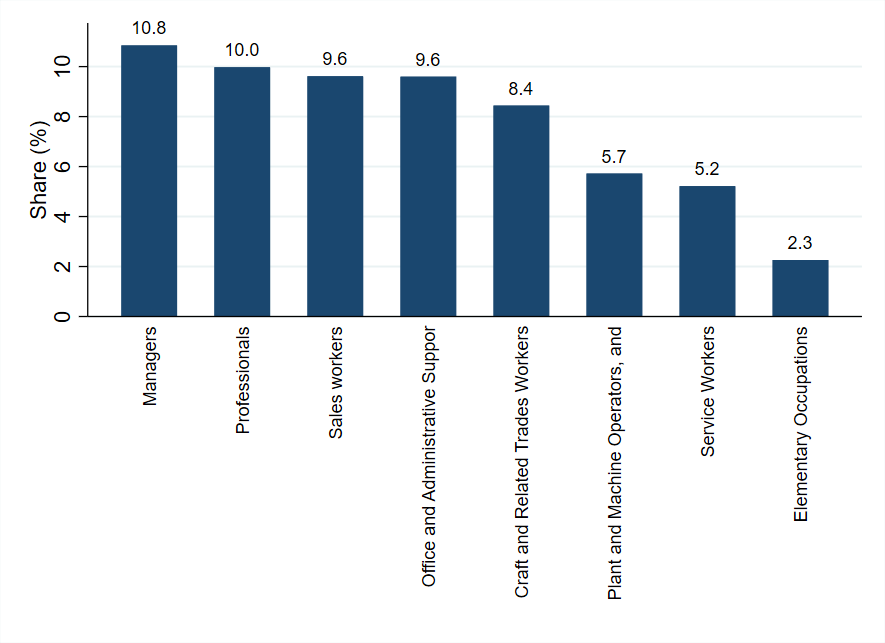}
    \caption{Occupations with decreased fulfilling-task share}
    \label{fig:less_fulfill_occ}
\end{subfigure}

\vspace{0.1cm}

\begin{subfigure}[b]{0.50\textwidth}
    \centering
    \includegraphics[width=\textwidth]{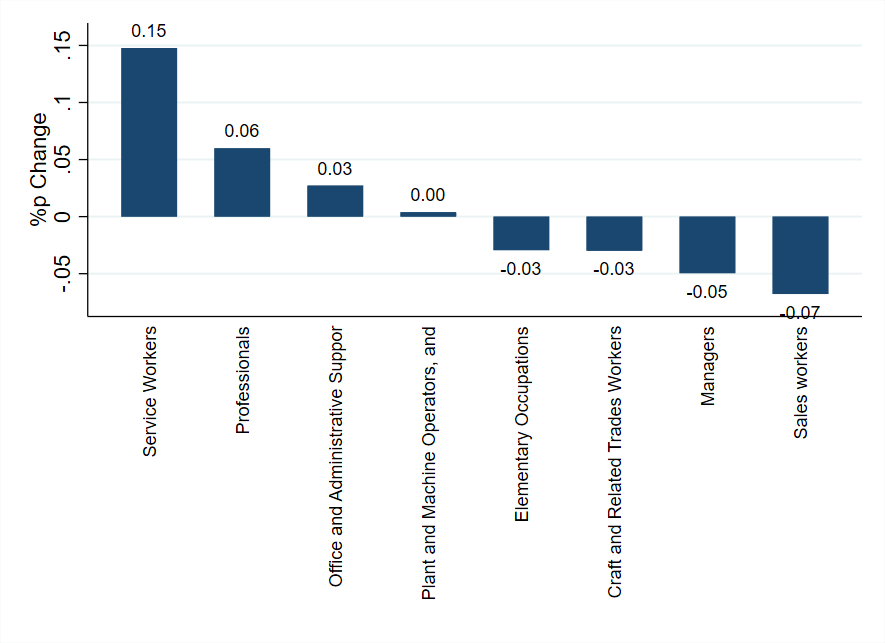}
    \caption{Net change in fulfilling-task share by occupation}
    \label{fig:diff_fulfill_occ}
\end{subfigure}

\caption{Changes in fulfilling-task share by occupation}
\label{fig:fulfill_comparison}
\end{figure}

\paragraph{Shifts in Fulfilling Work.}
Does this targeted efficiency shift the composition of work toward fulfilling activities? Figure \ref{fig:fulfill_comparison} summarizes the change in the share of time devoted to tasks rated as \textit{fulfilling} by respondents.

\subsection{Worker Satisfaction}

We estimate OLS regressions where the dependent variable is the change in job satisfaction (measured on a -5--5 scale). Table \ref{tab:satis_regressions} presents the results. We find a significant positive relationship between time savings and satisfaction. Crucially, this relationship holds even when controlling for output changes.

\input{outputs/satis_regressions.tex}

\section{A Conceptual Framework for Interpreting the Disconnect}

The empirical patterns documented above—substantial time savings, near-zero output correlation, and heterogeneous friction effects—can be understood through a simple task-based framework of labor supply. This framework clarifies the mechanisms underlying our findings and provides structure for welfare interpretation.

\paragraph{Environment.}
Workers allocate time ($T$) between production activities ($L_{work}$) and leisure ($L_{leisure}$). They derive utility from three sources: consumption $C$, leisure, and the fulfillment derived from their task portfolio ($M$):
\begin{equation}
U = u(C, L_{leisure}, M)
\end{equation}

Workers allocate total work hours across the tasks so that $L_{work}=\sum_i \tau_i \ell_i$ where $\tau_i$ is the efficiency of the worker at task $i$. Fulfillment depends on task composition. Let $\omega_i$ represent the intrinsic meaningfulness of task $i$ (measured in our survey by fulfillment ratings). Total fulfillment is the time-weighted average:
\begin{equation}
M = \sum_{i=1}^N \omega_i \ell_i
\end{equation}
Consumption is determined by wages: $C = Y$.

\paragraph{The AI Shock.}
GenAI alters the time cost of specific tasks. For a treated task $i$, GenAI increases the efficiency by $\alpha_i$, but may also introduce process frictions $\kappa_i \geq 0$ (prompting, verification, coordination overhead). The post-adoption time cost becomes:
\begin{equation}
\tau_i^{AI} = (1+\alpha_i)\tau_i - \kappa_i
\end{equation}

This shock creates \textit{latent productivity}—the potential for higher output at fixed effort. However, because labor supply is endogenous and workers value both leisure and fulfillment, they may re-optimize rather than expand output.

\paragraph{Three Mechanisms.}
This framework generates three patterns that map directly to our empirical findings:

\textbf{(1) The Leisure Channel:}
When GenAI increases task-level efficiency (high $\alpha$), workers face substitution and income effects. The substitution effect encourages more work; the income effect encourages more leisure. If labor supply elasticity is low—as is typical for salaried workers with fixed output quotas—the income effect dominates, and workers capture efficiency gains primarily as \textbf{leisure} rather than expanded output. This generates the near-zero correlation between time savings and output changes.

\textbf{(2) The Composition Channel:}
GenAI's impact varies across tasks with different $\alpha_i$ and $\omega_i$ (fulfillment). If AI primarily automates low-$\omega$ tasks (drudgery), workers reallocate time toward high-$\omega$ tasks (fulfilling work), raising welfare through the $M$ channel even if total output is flat. This generates the fulfillment shifts.

\textbf{(3) The Friction Channel:}
If frictions dominate efficiency gains ($\kappa_i > \alpha_i \tau_i$), the effective time cost \textit{increases}. Workers must spend more time to achieve the same output, or reduce output to preserve leisure. This explains the workers experiencing weakly increasing time and output declines.

\paragraph{Implications.}
This framework clarifies why efficiency gains need not translate to output growth in the short run. Workers exercise \textit{agency} over the productivity surplus, balancing output expansion against leisure and task quality. Realizing the full productivity boom will likely require institutional changes—performance incentives, job redesign, friction reduction—that shift the equilibrium from time-saving to output-expansion.

\section{Conclusion} \label{sec:conclusion}

This paper provides novel evidence on how GenAI affects time allocation, output, and welfare using a representative survey of Korean workers. We show that GenAI generates substantial efficiency gains—3.8\% time savings for users—but that these gains are currently captured as latent productivity. Rather than increasing output, workers utilize the efficiency to reduce work intensity (on-the-job leisure).

\newpage
\bibliographystyle{aer}
\bibliography{genai}

\newpage
\appendix

\section{International and Historical Comparison of GenAI Adoption} \label{app:comparison}

\begin{figure}[htp]
\centering
\begin{subfigure}{\textwidth}
  \centering
  \includegraphics[width=0.60\textwidth]{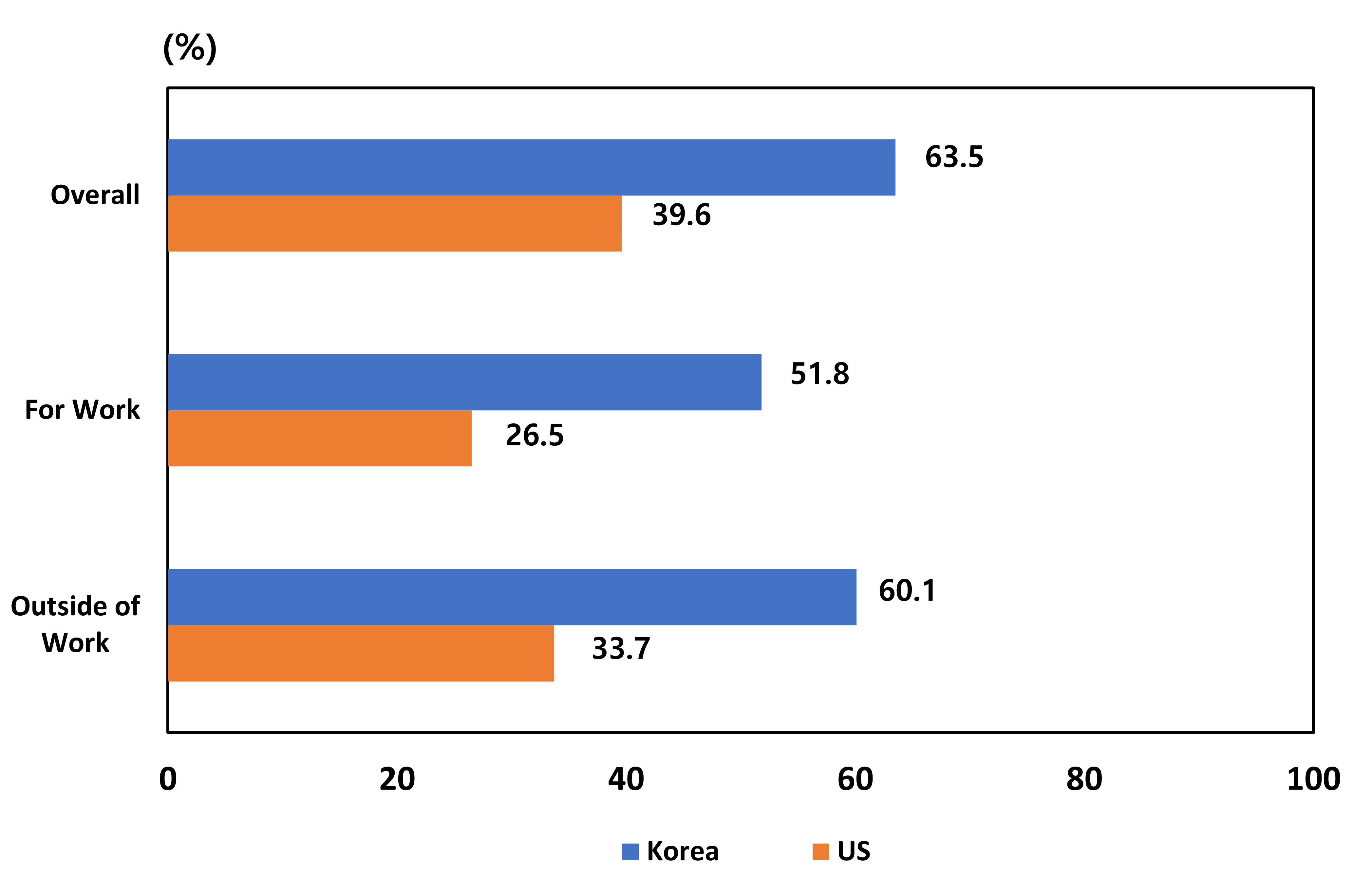}
  \caption{GenAI adoption: Korea vs. U.S. \citep{bick2025rapid}}
  \label{fig:adoption_us}
\end{subfigure}

\vspace{0.5cm}

\begin{subfigure}{\textwidth}
  \centering
  \includegraphics[width=0.60\textwidth]{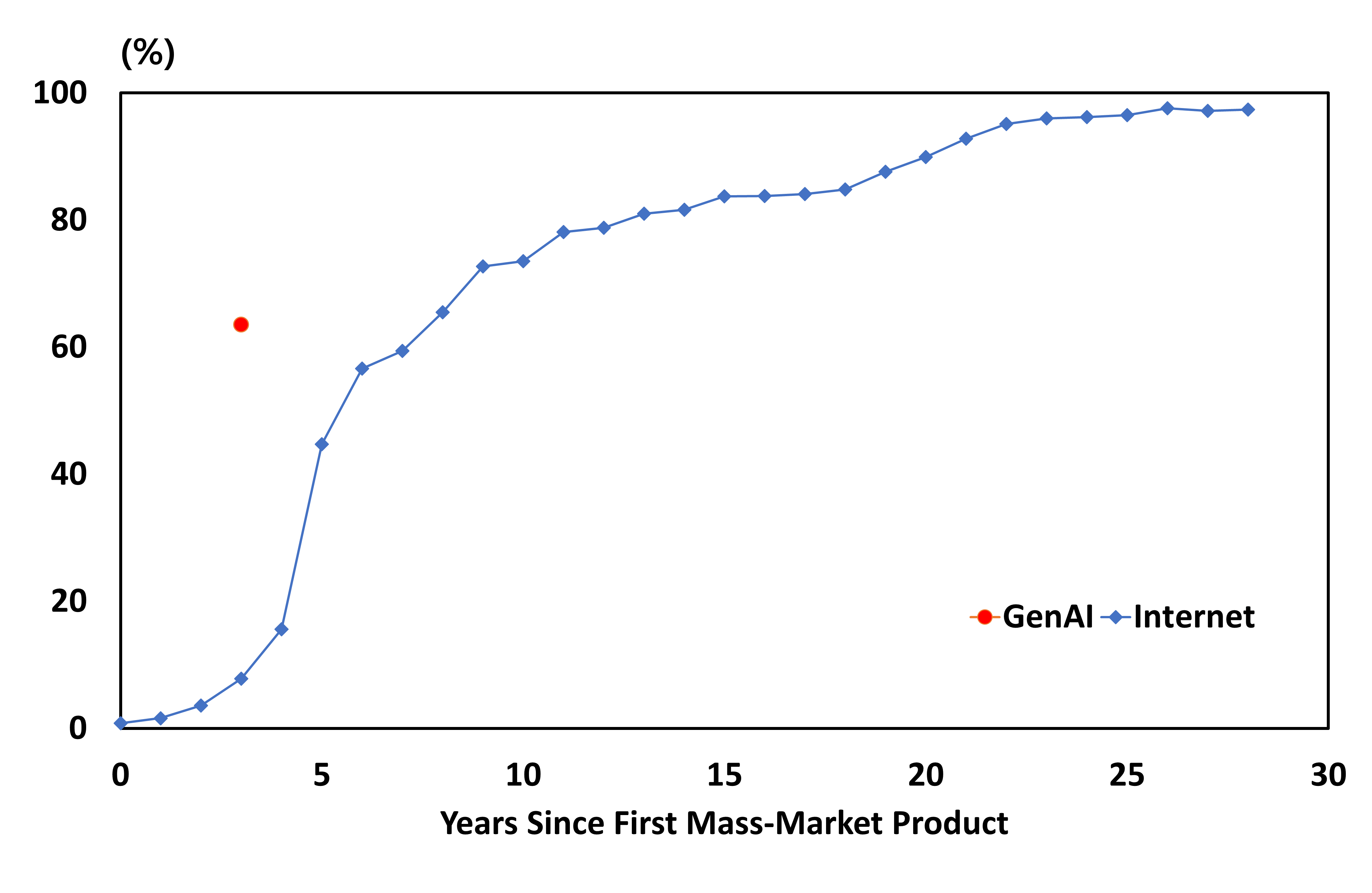}
  \caption{GenAI vs. internet adoption speed}
  \label{fig:adoption_internet}
\end{subfigure}

\caption{GenAI adoption in comparative perspective}
\label{fig:adoption_comparison}
\end{figure}

For context, Korea's GenAI adoption is roughly twice that of the U.S. across comparable categories, and the initial diffusion has been much faster than the historical rollout of the internet. Figure \ref{fig:adoption_us} compares Korean and U.S. adoption rates, and Figure \ref{fig:adoption_internet} contrasts GenAI diffusion with early internet adoption.

\end{document}

%% file: outputs/sample_summary.tex
\begin{tabular}{lrr}
\toprule
Category & Share & Type \\
\midrule
Sample size & 5,512 & Level \\
GenAI user (ever) & 51.8 & Share (\%) \\
GenAI user (last week) & 37.4 & Share (\%) \\
Gender: Male & 55.4 & Share (\%) \\
Gender: Female & 44.6 & Share (\%) \\
Age: 20--29 & 14.6 & Share (\%) \\
Age: 30--39 & 20.3 & Share (\%) \\
Age: 40--49 & 26.4 & Share (\%) \\
Age: 50--59 & 27.3 & Share (\%) \\
Age: 60--64 & 11.4 & Share (\%) \\
Education: 4yr College & 46.8 & Share (\%) \\
Education: Graduate & 9.7 & Share (\%) \\
\bottomrule
\end{tabular}

%% file: outputs/genai_main_results.tex
\begin{table}[htp]
\caption{Determinants of GenAI Adoption and Time Savings}
\label{tab:genai_main_results}
\centering
\footnotesize 
\renewcommand{\arraystretch}{1.1} 
\begin{tabular}{lcc}
\toprule
 & GenAI Adoption & Time Saved \\
 & (AME) & (\%) \\
\midrule
\multicolumn{3}{l}{\textit{\textbf{Demographics}}} \\
Male & 0.067*** (0.019) & 0.154 (0.641) \\
\multicolumn{3}{l}{\textit{Age (Ref: 50-59)}} \\
\hspace{3mm} 20-29 & 0.316*** (0.031) & 0.936 (1.142) \\
\hspace{3mm} 30-39 & 0.226*** (0.024) & -0.672 (0.651) \\
\hspace{3mm} 40-49 & 0.138*** (0.019) & 0.044 (0.591) \\
\hspace{3mm} 60+ & -0.021 (0.025) & 0.513 (0.787) \\
\midrule
\multicolumn{3}{l}{\textit{\textbf{Education} (Ref: High school)}} \\
Elementary or less & 0.169* (0.100) & 3.486 (2.609) \\
Middle school & 0.272* (0.160) & -2.841* (1.534) \\
2-3 year college & 0.025 (0.024) & 1.763* (1.017) \\
4-year college & 0.128*** (0.021) & 0.896 (0.568) \\
Master's degree & 0.178*** (0.034) & 0.360 (0.884) \\
Doctoral degree & 0.333*** (0.057) & 0.910 (1.419) \\
\midrule
\multicolumn{3}{l}{\textit{\textbf{Work-related characteristics}}} \\
Work motivation & 0.060*** (0.020) & 1.038* (0.590) \\
Experience (10y) & 0.013 (0.011) & -0.959*** (0.323) \\
Weekly hours (10h) & 0.023*** (0.007) & -0.060 (0.260) \\
Physical worker & -0.028* (0.017) & -0.246 (0.558) \\
Core task share & -0.049* (0.026) & -1.367 (0.870) \\
\midrule
\multicolumn{3}{l}{\textit{\textbf{Income and wealth}}} \\
Log income & -0.005 (0.009) & 0.414 (0.315) \\
Log wealth & 0.008 (0.005) & -0.152 (0.187) \\
\midrule
Observations & 4,943 & 1,835 \\
Adj. R$^2$ &  & 0.045 \\
\bottomrule
\end{tabular}

\vspace{0.1cm}
\begin{minipage}{0.95\linewidth} 
\scriptsize
\textit{Notes:} Robust standard errors in parentheses. Col 1: Logit Marginal Effects. Col 2: OLS on AI users only. Controls include employment status, firm size, occupation, and industry FE. * $p<0.1$, ** $p<0.05$, *** $p<0.01$.
\end{minipage}
\end{table}

%% file: outputs/output_regressions.tex
\begin{table}[htp]
\caption{Determinants of Output Change: Worker vs. Task Level}
\label{tab:output_regressions}
\centering
\footnotesize 
\renewcommand{\arraystretch}{1.1} 
\begin{tabular}{lccc}
\toprule
 & Worker-Level & Task-Level & Task-Level + FE \\
 & (1) & (2) & (3) \\
\midrule
\textbf{Time Saved (\%)} & 0.000 (0.162) &  &  \\
\textbf{Task-Level Time Saved (\%)} &  & -0.223 (0.867) & -0.183 (0.935) \\
\midrule
\multicolumn{4}{l}{\textit{\textbf{Demographics}}} \\
Male & -3.297*** (1.238) & -3.055** (1.279) & -2.999** (1.291) \\
\multicolumn{4}{l}{\textit{Age (Ref: 50s)}} \\
\hspace{3mm} 20-29 & 5.926*** (2.199) & 4.554** (2.151) & 4.831** (2.332) \\
\hspace{3mm} 30-39 & 4.994*** (1.621) & 5.039*** (1.584) & 5.289*** (1.796) \\
\hspace{3mm} 40-49 & 3.491** (1.364) & 3.447*** (1.269) & 3.926*** (1.336) \\
\hspace{3mm} 60+ & 2.802 (1.825) & 2.050 (1.752) & 1.822 (1.908) \\
\midrule
\multicolumn{4}{l}{\textit{\textbf{Education} (Ref: High school)}} \\
Elem. or less & -0.809 (5.841) & -0.202 (5.150) & 0.211 (5.231) \\
Middle school & -0.328 (4.177) & -2.021 (5.242) & -4.488 (4.966) \\
2-3y college & 1.372 (1.650) & 1.402 (1.511) & 1.304 (1.530) \\
4-year college & 3.157** (1.528) & 2.698* (1.455) & 2.878* (1.492) \\
Master's & 2.712 (2.303) & 2.881 (2.277) & 3.818 (2.451) \\
PhD & 0.251 (3.189) & -1.981 (3.901) & -0.347 (3.847) \\
\midrule
\multicolumn{4}{l}{\textit{\textbf{Work Characteristics}}} \\
Work motivation & -3.240** (1.339) & -4.156*** (1.225) & -4.336*** (1.455) \\
Experience (10y) & 0.775 (0.725) & 0.033 (0.857) & 0.096 (0.975) \\
Weekly hours (10h) & -0.489 (0.495) & -0.563 (0.469) & -0.421 (0.526) \\
Physical worker & -2.639** (1.139) & -2.605** (1.313) & -2.560* (1.502) \\
Core task ratio & 0.656 (2.226) & 0.833 (2.153) & 0.980 (2.012) \\
\midrule
\multicolumn{4}{l}{\textit{\textbf{Income and wealth}}} \\
Log income & -0.042 (0.452) & -0.027 (0.447) & -0.242 (0.481) \\
Log wealth & 0.093 (0.330) & 0.019 (0.371) & 0.089 (0.376) \\
\midrule
Observations & 1,835 & 5,769 & 5,769 \\
Adj. $R^2$ & 0.033 & 0.044 & 0.020 \\
\bottomrule
\end{tabular}

\vspace{0.1cm}
\begin{minipage}{0.95\linewidth} 
\scriptsize
\textit{Notes:} Robust standard errors in parentheses (clustered at the worker level for Models 2-3). Model 1 is a worker-level OLS. Models 2 and 3 are task-level regressions (workers report AI usage for multiple specific tasks). Model 3 includes Task Content Fixed Effects (e.g., Writing, Coding, Analysis). * $p<0.1$, ** $p<0.05$, *** $p<0.01$.
\end{minipage}
\end{table}

%% file: outputs/task_concentration_tables.tex
\providecommand{\toprule}{}\providecommand{\midrule}{}\providecommand{\bottomrule}{}
\begin{table}[htp]
  \centering
  \caption{Time-savings concentration by occupation}
  \label{tab:task_concentration}
  \begin{tabular}{lrrr}
    \toprule
    Occupation (aggregated) & Mean top-task share & Share(top$\ge$0.8) & N \\
    \midrule
    Sales \& customer-service managers & 0.45 & 0.10 & 20 \\
    IT professionals/technicians & 0.64 & 0.37 & 65 \\
    Health/social/religious professionals & 0.80 & 0.64 & 26 \\
    Education professionals & 0.61 & 0.36 & 64 \\
    Legal/public admin professionals & 0.69 & 0.39 & 36 \\
    Business/finance professionals & 0.61 & 0.27 & 41 \\
    Culture/arts/sports professionals & 0.84 & 0.65 & 43 \\
    Science/engineering professionals & 0.66 & 0.35 & 72 \\
    Clerical (business/accounting) & 0.77 & 0.56 & 43 \\
    Finance clerks & 0.71 & 0.47 & 33 \\
    Admin/secretarial support & 0.65 & 0.34 & 38 \\
    Manufacturing process workers & 0.64 & 0.42 & 23 \\
    Skilled trades & 0.63 & 0.30 & 29 \\
    Service workers & 0.67 & 0.38 & 98 \\
    \bottomrule
  \end{tabular}
\end{table}

\begin{table}[htp]
  \centering
  \caption{Time-savings concentration by industry}
  \label{tab:task-concentration-industry}
  \begin{tabular}{lrrr}
    \toprule
    Industry (aggregated) & Mean top-task share & Share(top$\ge$0.8) & N \\
    \midrule
    Agriculture/forestry/fishery & 0.75 & 0.51 & 24 \\
    Mining/utilities & 0.71 & 0.45 & 174 \\
    Manufacturing (durables) & 0.75 & 0.58 & 22 \\
    Construction & 0.75 & 0.51 & 39 \\
    Wholesale/retail & 0.65 & 0.35 & 66 \\
    Information/communication & 0.69 & 0.44 & 82 \\
    Finance/insurance & 0.67 & 0.40 & 68 \\
    Professional/science/tech services & 0.64 & 0.35 & 105 \\
    Public administration/defense & 0.73 & 0.50 & 50 \\
    Education services & 0.65 & 0.38 & 109 \\
    Health/social work & 0.74 & 0.53 & 59 \\
    Arts/sports/other services & 0.84 & 0.64 & 43 \\
    \bottomrule
  \end{tabular}
\end{table}

%% file: outputs/satis_regressions.tex
\begin{table}[htp]
\caption{Determinants of Satisfaction Change}
\label{tab:satis_regressions}
\centering
\footnotesize 
\renewcommand{\arraystretch}{1.1} 
\begin{tabular}{lc}
\toprule
 & Satisfaction Change \\
 & (1) \\
\midrule
\textbf{Time Saved (\%)} & 0.039*** (0.008) \\
\textbf{Output Change (\%)} & 0.007*** (0.002) \\
\midrule
\multicolumn{2}{l}{\textit{\textbf{Demographics}}} \\
Male & -0.035 (0.096) \\
\multicolumn{2}{l}{\textit{Age (Ref: 50s)}} \\
\hspace{3mm} 20-29 & 0.384** (0.176) \\
\hspace{3mm} 30-39 & 0.265* (0.144) \\
\hspace{3mm} 40-49 & -0.011 (0.125) \\
\hspace{3mm} 60+ & 0.015 (0.179) \\
\midrule
\multicolumn{2}{l}{\textit{\textbf{Education} (Ref: High school)}} \\
Elem. or less & 0.854 (0.528) \\
Middle school & -0.381 (0.579) \\
2-3y college & 0.147 (0.165) \\
4-year college & 0.208 (0.144) \\
Master's & 0.250 (0.184) \\
PhD & 0.152 (0.302) \\
\midrule
\multicolumn{2}{l}{\textit{\textbf{Work Characteristics}}} \\
Work motivation & 0.174 (0.112) \\
Experience (10y) & -0.015 (0.065) \\
Weekly hours (10h) & 0.055 (0.041) \\
Physical worker & 0.444*** (0.090) \\
Core task ratio & 0.035 (0.154) \\
\midrule
\multicolumn{2}{l}{\textit{\textbf{Income and wealth}}} \\
Log income & 0.089* (0.047) \\
Log wealth & 0.053* (0.028) \\
\midrule
Observations & 1,835 \\
Adj. $R^2$ & 0.120 \\
\bottomrule
\end{tabular}

\vspace{0.1cm}
\begin{minipage}{0.5\linewidth} 
\scriptsize
\textit{Notes:} Robust standard errors in parentheses. OLS regression estimated on AI users only. Dependent variable is the reported change in job satisfaction due to GenAI usage (scale: -5 to +5). Main independent variables are percentage time savings and output change. * $p<0.1$, ** $p<0.05$, *** $p<0.01$.
\end{minipage}
\end{table}